\documentclass[]{aastex631}

\usepackage{amsmath,amssymb, amsthm,amstext}
\usepackage{natbib}
\usepackage{graphicx}
\usepackage{color}
\usepackage{array, enumerate}
\usepackage{bm}
\usepackage{multirow}
\usepackage{braket}
\usepackage{txfonts}

\begin{document}
\title{Measuring Mass Transfer Rates in Coalescing Neutron Star--White Dwarf Binaries \\ with Deci-Hz Gravitational-wave Detectors}
\author[0000-0002-0151-3794]{Zhenwei Lyu}
\email{zwlyu@pku.edu.cn}
\affiliation{Kavli Institute for Astronomy and Astrophysics, Peking University, Beijing 100871, China}
\author[0000-0002-1334-8853]{Lijing Shao}
\email{lshao@pku.edu.cn}
\affiliation{Kavli Institute for Astronomy and Astrophysics, Peking University, Beijing 100871, China}
\affiliation{National Astronomical Observatories, Chinese Academy of Sciences, Beijing 100012, China}

\begin{abstract}
Coalescing neutron star--white dwarf (NS-WD) binaries are among the primary targets for upcoming space-borne gravitational wave (GW) detectors such as LISA, TaiJi, TianQin, etc. During close interaction, these binaries undergo mass transfer, emitting simultaneous X-rays and GWs. This offers a unique opportunity to measure mass transfer rates and study compact binary evolution. To analyze mass transfer rates, we employ the TaylorF2 frequency domain waveform model within the stationary phase approximation (SPA). Through this approach, we derive the GW phase induced during the mass transfer phase and perform Markov Chain Monte Carlo (MCMC) simulations to estimate the minimal detectable mass transfer rate given specific signal-to-noise ratios (SNRs). Our results suggest that for a NS-WD binary with a $0.5 \rm M_\odot$ white dwarf companion, we could measure mass transfer rates down to $10^{-7}\rm M_\odot , {\rm yr}^{-1}$ at SNR=20 and $10^{-9}\rm M_\odot , {\rm yr}^{-1}$ at SNR=1000. This measurement holds significance for studying compact binary evolution involving mass transfer and has potential applications in forecasting tidal disruption events.




\end{abstract} 

\section{Introduction} 
The detection of gravitational wave (GW) emissions from closely orbiting compact binaries, comprising stellar-mass objects such as white dwarfs (WDs), neutron stars (NSs), or black holes (BHs), stands as a crucial scientific objective for space-borne GW detectors \citep{Kupfer_2018,Tauris_2018,Amaro_Seoane_2023,Chen_2020, Bayle:2022hvs,Liu_2022,Li_2023}. These binaries can be broadly categorized as detached or interacting, with our emphasis of this {\it Letter} on exploring the mass transfer (MT) effects within interacting binaries involving WDs. Specifically, we focus on ultra-compact X-ray binaries (UCXBs), typically WD+NS/BH systems, where simultaneous emission of electromagnetic radiation and GWs takes place \citep{Nelemans_2010,kaltenborn_2022,Kang_2023}.\\
 
In interacting binary systems, significant MT occurs when the donor's radius exceeds its Roche lobe radius, resulting in a mass flow from the donor to the accretor. This study particularly considers a WD as the donor ($m_2$), with the accretor ($m_1$) being a NS or BH and $m_1 > m_2$ is assumed. This choice is influenced by the observation that a closer orbit not only leads to a higher MT rate and higher signal-to-noise ratios (SNRs) but also generates GW radiation within the frequency band $(10^{-3}, 10^{-1})$\,Hz, which corresponds to the sensitive range for space-borne GW detectors \citep{Wang_2023}. Notable projects in this field include the Laser Interferometer Space Antenna (LISA) led by the European Space Agency (ESA) and the National Aeronautics and Space Administration \citep[NASA;][]{babak2021lisa,amaroseoane2017laser,Robson_2019}, two Chinese programs—TaiJi \citep{Gong_2021,Ruan_2020,Hu_2017,Liu_2023} and TianQin \citep{Gong_2021,Luo_2016,Luo_2020,Mei_2020}—as well as the Deci-hertz Interferometer Gravitational wave Observatory (DECIGO)---and B-Decigo, a scientific pathfinder for DECIGO---proposed by the Japanese \citep{Isoyama_2018,kawamura2020current}. Lunar detectors are also part of this exploration \citep{Jani_2021,Harms_2021,Junlang_2023,cozzumbo2023opportunities}. As illustrated in Fig.~\ref{fig:sensitivity}, TianQin and B-Decigo exhibit higher sensitivity in the frequency band $(10^{-2}, 10^{-1})$\,Hz, potentially offering a more stringent constraint on MT effects.\\

In recent years, many UCXBs have been discovered, sparking debates on the long-term stability and mass distribution of WD donors. The analysis of stable Roche lobe overflow suggests that WD masses $m_2<1.25\rm M_\odot$ are required for a NS accretor of mass $m_1=1.4\rm M_\odot$ \citep{Yu_2021}. Formation channel studies indicate that the initial mass distribution of WD donors in UCXBs can be classified into two groups. The majority are helium WDs with masses below about $0.35 \rm M_\odot$, and the secondary group comprises hybrid WDs with higher masses ranging from $0.35$ to $0.65 \rm M_\odot$ \citep{Nelemans_2010_xray}. The investigation of optical, ultraviolet, and X-ray spectroscopy from UCXBs reveals the overabundance of carbon and oxygen lines, a phenomenon that has been extensively discussed \citep{Juett_2001,Schulz_2001,Homer_2002,Juett_2003,Nelemans_2004,Werner_2006,Nelemans_2006}. This observation implies the presence of Carbon/Oxygen (C/O) disks in these systems, and hybrid WDs with masses less than $0.45\rm M_\odot$ are believed to be necessary to account for the existence of C/O disks \citep{yungelson2002formation,Nelemans_2010,Heinke_2013}. Additionally, there exist studies discussing the presence of UCXBs with BHs as accretors \citep{Sberna_2021,Chen_2023}. 
Without loss of generality, we assume that mass transfer commences when the radius of the WD fills its effective Roche lobe and terminates at the tidal disruption radius, as discussed in Appendix~\ref{sec:continuous_MT}. Following this, four examples of WD-NS/BH systems have been devised to examine the MT rate (MTR) in this {\it Letter}, as detailed in Table~\ref{tab:cases}. \\

MTR is a crucial factor in studying the accretion mechanism, a topic extensively discussed in recent studies of binary evolution and compact object formations \citep{Sengar_2017,Yu_2021,Chen_2022}. However, current estimations of MTR are extremely challenging and heavily reliant on specific models \citep{yungelson2002formation,Heinke_2013,Yu_2021,Chen_2022}. As we show, GWs offer a model-independent and direct method to measure the MTR. Successfully measuring the MTR in NS-WD systems will advance our understanding of accretion mechanisms involving WDs, as well as the physics of ionized accretion disks and jets, which are essential for electromagnetic observations. Furthermore, the detection of MTR will aid in forecasting of tidal disruption events and facilitate multi-messenger observations \citep{Lodato_2010,Abbott_2017,Dai_2018,Maguire_2020}. \\

This {\it Letter} is structured as follows: Firstly, in Sec. \ref{sec:waveform}, we construct the waveform model incorporating MT effects based on the stationary phase approximation (SPA). Additionally, we explore the degeneracy and potential methods to resolve it. In Sec. \ref{sec:pe}, we compare the MT-induced correction to the post-Newtonian (PN) phase of GWs at various orders before engaging in Bayesian parameter estimation. Finally, we present a conclusion and further discussion in Sec. \ref{sec:conclusion}. \\

Throughout, we denote $m_2$ as the mass of the WD, and $m_1$ as the mass of the companion (NS/BH), where $m_1 > m_2$. We adopt the conversion $c=G=1$, indicating that mass, length, and time share the same dimension. The true physical dimensions can be easily restored using the conversion: $1\,\rm M_\odot = 1.4766\,\rm km = 4.9255 \times 10^{-6}\,\rm s$.

\section{Waveform Model}\label{sec:waveform}
 We specifically examine MT during the late inspiral stage of a binary, utilizing the analytically constructed frequency domain waveform model, TaylorF2, within the framework of stationary phase approximation \citep[SPA;][]{Damour_2001,Arun_2005,Buonanno_2009}. The TaylorF2 waveform is formulated as
\begin{align}
    \tilde{h}_+(f) = \mathcal{A} \left(\frac{1+\cos^2{\iota}}{2} \right)e^{i\,\psi(f)} \,,\quad
    \tilde{h}_\times(f) = i\,\mathcal{A} \cos{\iota}\,e^{i\, \psi(f)} \,,        
\end{align}
while $\iota$ is the inclination angle between the orbital angular momentum $\bm{L}$ and the line-of-sight. In this study we set $\iota=0$, corresponding to an optimally oriented source. The coefficients $\mathcal{A}$ and $\psi$ are given by
\begin{align}
    \mathcal{A} = \pi^{-2/3}\left(\frac{5}{24}\right)^{1/2} \frac{\mathcal{M}_c^{\,5/6}}{d}\,f^{\,-7/6}\,,\quad
    \psi(f) = \psi_{\rm pp} + \psi_{\rm tide} + \psi_{\rm spin} + \psi_{\rm MT}\,,\label{eq:psiTot}
\end{align}
where $\mathcal{M}_c$ is the chirp mass, and $d$ is the luminosity distance of the system. The terms $\psi_{\rm pp}, \psi_{\rm spin}, \psi_{\rm tide}$, and $\psi_{\rm MT}$ represent the SPA phase in the point particle limit (spin-independent part), spin-induced phase, tidal phase,  and the phase induced by the MT, respectively. These terms are discussed explicitly in Appendix~\ref{sec:waveform_model} .\\

Given the challenge posed by the mass-distance degeneracy in the amplitude coefficient $\mathcal{A}$ , achieving accurate measurements of binary masses through amplitude, and subsequently determining MTR, is difficult. As a result, we currently ignore higher PN order corrections to the amplitude. Alternatively, we shift our attention to the phase, which, as demonstrated in current LIGO data analysis \citep{Pan_2020}, can be precisely measured to $\mathcal{O}(1)$, and we anticipate that the phase resolution of Deci-Hz GWs will achieve a similar level of accuracy.\\

\subsection{MT-Induced Phase}
Due to the MT, the adiabatic approximation is no longer valid. We must employ a modified balance equation,
\begin{align}
    \frac{d E(v;m_i(v))}{d t} + \mathcal{F}(v;m_i(v)) + \dot{M} = 0\,,
\end{align}
where $m_i(v)$ is the mass of the $i$-th object, and $\dot{M}$ is the derivative of the total mass with respect to time. $E(v;m_i(v))$ and $\mathcal{F}(v;m_i(v))$ are the orbital energy and the GW flux at infinity, respectively. With the SPA approach, we derived the MT-induced phase in Appendix~\ref{subsec:MT_phase}. The total mass conserved term reads 
\begin{align}\label{eq:phiMT1}
    \psi_{\rm MT} &= -\frac{5\,\xi\,\dot{m}_2\,\delta v^{\rm I}}{4096\,\eta^3\,v^{14}}\,,   
\end{align}
where $\xi$ is a function of $m_1$ and $m_2$ as defined in Appendix~\ref{subsec:MT_phase}, $\eta$ denotes the symmetric mass ratio, and $\dot{m}_2$ is the MTR of the donor. Given that the waveform is in the frequency domain, $\delta v^{\rm I}=v-v_i$ where $v_i$ is the initial value of $v$. This term is proportional to $v^{-13}=v^{-5}\,v^{-8}$ which is at the $-4$PN order, consistent with the variation of the gravitational constant \citep{Damour_1988, Yunes_2010,Tahura_2018, Wang:2022yxb}. Concerning $\dot{m}_2$, the dimensionless quantity $G \dot{m}_2/c^3 = -1.56\times 10^{-21}$ is remarkably small, corresponding to a MTR of $-10^{-8} \, \rm M_\odot \, {\rm yr}^{-1}$ for $m_2$. Because the velocity $v$ is of order $\mathcal{O}(10^{-2})$ in the denominator, the MT-induced phase $\psi_{\rm MT}\approx10$ rad for the case NSWD2 in Table~\ref{tab:cases}, indicating that a MTR of $-10^{-8} \, \rm M_\odot \, {\rm yr}^{-1}$ is measurable. For convenient estimation of the small MTR, we introduce the power index $\alpha = \log_{10}(-\dot{m}_2 / (\rm M_\odot \, {\rm yr}^{-1}))$. \\

Mathematically, the deviation above of MT-induced phase is applicable to inspiraling point-particle binaries. However, MT effects are only exhibited in the close interaction of extended objects, where tidal and spin-induced effects are expected to be significant, particularly for WDs. It is essential to account for these effects in the total phase. Furthermore, the inclusion of higher PN order terms are crucial for partially breaking the degeneracies along masses and MTR, as discussed in Appendix~\ref{subsec:degeneracy}. We incorporate higher PN order terms up to $3.5$PN for $\psi_{\rm pp}$ and $\psi_{\rm spin}$, and tidal effects for $\psi_{\rm tide}$ up to $7.5$PN, as discussed in Appendix~\ref{subsec:tidal_spin}. Moreover, we make the assumption of total mass conservation ($\dot{M}=0$) due to the complete degeneracy between $\beta$ and $\alpha$ as shown in Eq.~(\ref{eq:phiMT2}). \\

\section{Parameter Estimation}\label{sec:pe}
In Appendix~\ref{sec:continuous_MT}, we provide a concise discussion on continuous MT effect. Given the differences in the time duration from the initiation of Roche lobe overflow to tidal disruption across different scenarios (see Table~\ref{tab:cases}) and considering the limited operation time of GW detectors, a common practice is to set a fixed observing duration of 4 years. Consequently, the initial frequency $f_{\rm i}$ in the phase comparison and MCMC simulation is defined as 4 years before the tidal disruption; the Bayesian inference is discussed in Appendix~\ref{subsec:Bayes}.   \\

we generate four binary scenarios with configurations specified in Table~\ref{tab:cases}, with various mass for the WD donor and an accretor that is either a NS or a BH. Initially, we assess the measurability of the MTR by comparing phases. Subsequently, to better evaluate the uncertainties in MTR measurements across different scenarios, we conduct MCMC simulations.\\

\renewcommand{\arraystretch}{1.4}
\begin{table*}[tb]
\caption{Representative configurations of binary scenarios involving $m_1$ as the mass of the accretor and $m_2$ as the mass of the donor. $f_{\rm R}$ denotes the GW frequency at the point where the Roche lobe radius reaches the WD radius, and $f_{\rm t}$ represents the GW frequency corresponding to the tidal disruption radius. Furthermore, $f_{\rm i}$ corresponds to the GW frequency four years before tidal disruption. The variable $T_{\rm MT}$ represents the time duration from $f_{\rm R}$ to $f_{\rm t}$. Subsequently, SNRs are computed at a luminosity distance of $d_{\rm L}=20\;\rm kpc$ under varying sensitivities shown in Fig.~\ref{fig:sensitivity}. The last two columns present the lowest achievable $\alpha$ in the injection, along with the associated evaluation with 1-$\sigma$ error when SNR$=20$. Regarding the NSWD1 case, it demonstrates the impossibility of resolving the degeneracy and achieving a constraint on $\alpha$.
}
\centering
\setlength{\tabcolsep}{3pt}
\begin{tabular}{c|c c | c c c c|c c c c c |c c}
\hline
\hline
\multirow{2}{1.5em}{} & \multirow{2}{2.5em}{$m_1 (\rm M_\odot)\;$} & \multirow{2}{3em}{$\;m_2 (\rm M_\odot)$} & \multirow{2}{1em}{$f_{\rm R}\, (\rm Hz)$} & \multirow{2}{1em}{$f_{\rm i}\, (\rm Hz)$} & \multirow{2}{1em}{$f_{\rm t}\, (\rm Hz)$} & \multirow{2}{3.5em}{$T_{\rm MT} \left(\rm yrs\right)\;$} & \multicolumn{5}{c|}{$\rm SNR\left(\it d_{\rm L}\rm =20\, kpc\right)$} & \multicolumn{2}{c}{$\alpha$}
 \\
 & & & & & &  & Lunar & LISA & TaiJi & TianQin & B-Decigo & injection & evaluation\\
\hline
NSWD1 & 1.4  & 0.1 & 0.0049645& 0.0163651& 0.0163677& 224000 & 0.03 & 40 & 151 & 90 & 56 & -- & -- \\[0.5ex]
\hline
NSWD2 & 1.4 & 0.5 & 0.02611 & 0.09062 & 0.09774 & 580 & 37 & 179 & 600 & 1300 & 27300 & $-7.1$ & $-7.2^{+0.17}_{-0.16}$ \\[0.5ex]
\hline
NSWD3 & 1.4 & 1.0 & 0.09570 & 0.13316 & 0.39028 & 10 & 1270 & 274 & 917 & 3470 & 435000 & $-7.5$ & $-7.65^{+0.22}_{-0.65}$ \\[0.5ex]
\hline
BHWD & 20  & 0.5 & 0.02698 & 0.06557 & 0.08494 & 82 & 115 & 1230 & 4210 & 7280 & 95100 & $-7.7$ & $-7.63^{+0.14}_{-0.25}$ \\
\hline
\end{tabular}
\label{tab:cases}
\end{table*}

\subsection{Comparison of Phases}
\begin{figure*}[ht!]
    \centering
    \includegraphics[trim={0pt 0pt 0pt 0pt},clip,width=0.95\linewidth]{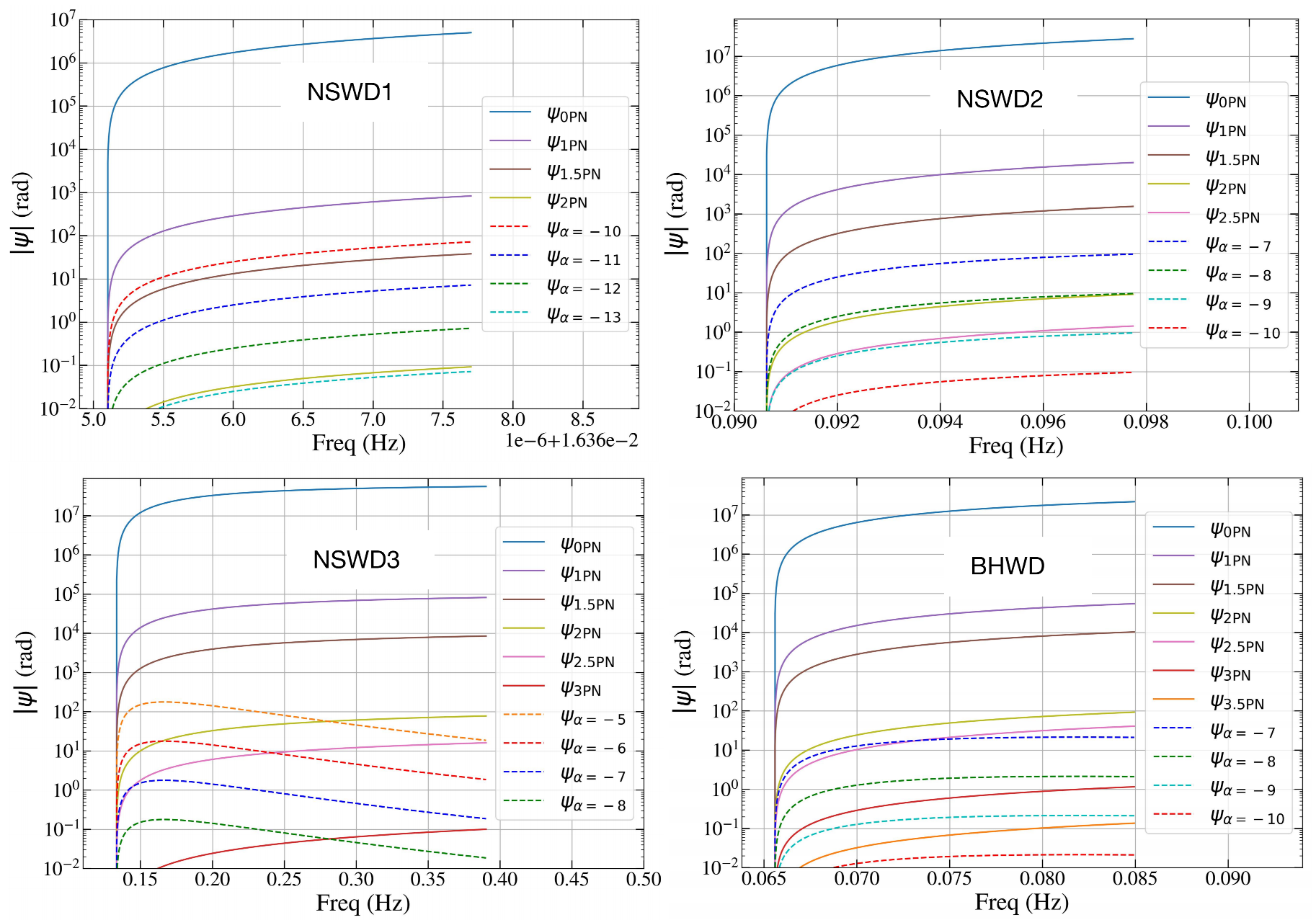}
    \caption{The SPA phase $|\phi|$ plotted against frequency $f$ for four binary scenarios in Table~\ref{tab:cases}. The initial frequency corresponds to 4 years before tidal disruption as listed in the fifth column of Table~\ref{tab:cases}, while the tidal disruption frequency is listed in the sixth column. The power index of MTR from $m_2$ is denoted by $\alpha = \log_{10}(-\dot{m}_2 / (\rm M_\odot \, {\rm yr}^{-1}))$. }
    \label{fig:phi}
\end{figure*}

We illustrate the phases arising from MT effects and compare them to PN orders from the point-particle limit, as depicted in Fig.~\ref{fig:phi}. The MT-induced phase reaches $\mathcal{O}(1)$ when $\alpha$ is set to $-12$, $-9$, $-7$ and $-8$ for NSWD1, NSWD2, NSWD3 and BHWD cases,  respectively. In principle, approximately these are the lowest $\alpha$ that we can measure.\\

When comparing the MT-induced phase resulting from different values of $\alpha$, an important feature is the linear dependence of the MT-induced phase on the power index $\alpha$ and, consequently, on $\dot{m}_2$, consistent with Eq.~(\ref{eq:phiMT1}). Moreover, in the comparison of the phase across three NS-WD scenarios, one finds that a slower WD mass leads to a greater MT-induced phase. Another noteworthy observation is the comparison between NSWD2 and BHWD scenarios, revealing that a lower accretor mass leads to a relatively higher MT-induced phase. \\

When comparing with PN orders, it is evident that the phase induced by the first few leading PN orders from $\psi_{\rm pp}$ is significantly greater than the detection limit of $\mathcal{O}(1)$. The inclusion of additional higher-order terms in the total phase is necessary and, in principle, will help to break the degeneracy along $\mathcal{M}_c$, $\eta$ and $\alpha$ as discussed in Appendix~\ref{subsec:degeneracy}.\\

\subsection{Simulation Results}
We examine a four-year evolution of binary systems terminating at the tidal disruption frequency (see Table~\ref{tab:cases}). To draw a general conclusion that is independent of specific detectors, a binary system is set up at a distance where its SNR is $20$. In the last two columns of Table~\ref{tab:cases}, the injected and estimated values of the lowest detectable $\alpha$ are presented. The simulation shows notable variation in the MT-induced phase for the lowest detectable $\alpha$ across scenarios. This is attributed to the frequency range covered over four years. A higher WD mass, with a lower radius and higher GW radiation, leads to a broader frequency range in 4 years. This might assist in resolving the degeneracy and offer a more precise constraint on MTR. In the scenario of NSWD1, representing the smallest considered WD mass, it suggests the challenge of resolving the degeneracy between masses and $\alpha$. This difficulty primarily stems from the WD's significantly large radius and the exceptionally minor frequency evolution over a four-year span, ranging from $f_{\rm i}=0.0163651$ Hz to $f_{\rm t}=0.0163677$ Hz. \\

\begin{figure*}[t!]
	\centering
    \includegraphics[trim={2cm 0pt 2cm 0pt},clip,width=0.98\linewidth]{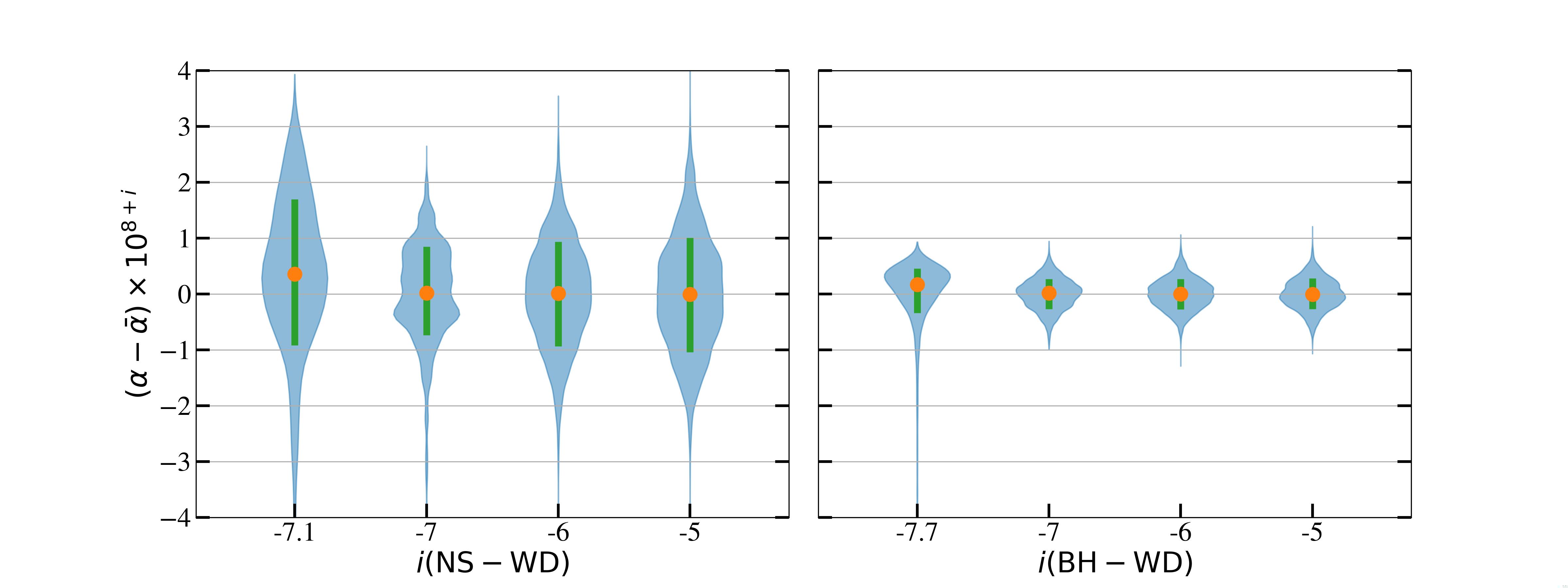}
    \caption{The violin plot illustrates NSWD2 (left) and BHWD (right) cases at an SNR of 20, depicting 1-$\sigma$ uncertainties with green bars and representing mean values with orange dots. The x-axis labeled with $i$ corresponds to the injected value of $\alpha$. Notably, the plot reveals that the 1-$\sigma$ error associated with $\alpha$ exhibits a proportional decrease of $10^{\,\delta\alpha}$, as $\alpha$ varies by $\delta\alpha$.
}\label{fig:violin}
\end{figure*}

With a given SNR, the resolvable phase $\Delta \psi_{\rm MT}$ remains nearly constant. As the parameter $\alpha$ increases by $\delta\alpha$, there is a proportional decrease of $10^{\,\delta\alpha}$ in the 1-$\sigma$ error bound, because $\Delta \psi_{\rm MT} \propto \Delta\dot{m}_2 \propto \Delta(-10^{\alpha})\propto 10^{\alpha}\,\Delta\alpha$. Our simulations with a relative higher $\alpha$ provide evidence for this claim, as shown in Fig.~\ref{fig:violin}. \\

\begin{figure*}[ht]
	\centering
    \includegraphics[width=0.95\textwidth]{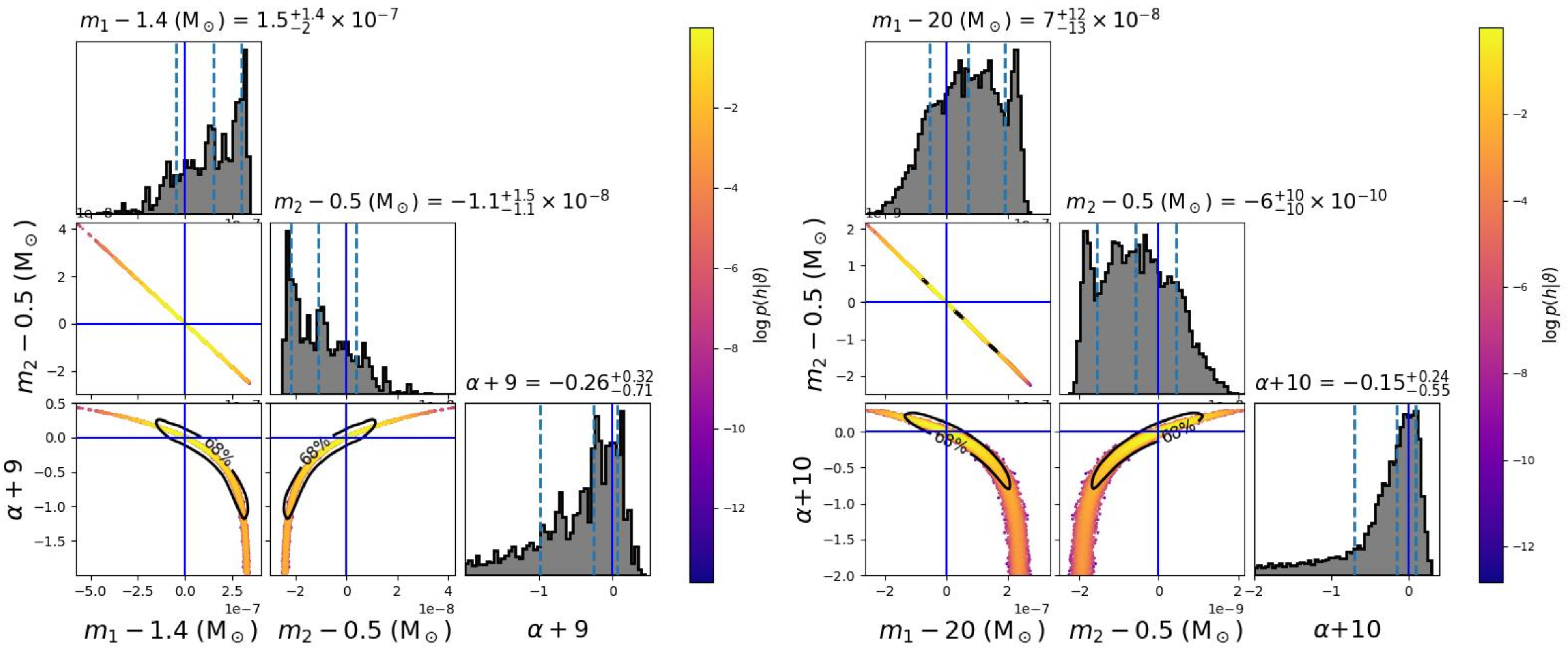}
    \caption{The posterior distributions of $m_1$, $m_2$, and $\alpha$ demonstrate the measurement of scenarios with nearly minimally detectable $\alpha$ when the systems are at $d_{\rm L}=0.73 \, \rm kpc$ in both NSWD2 (left panel, SNR $\approx 1000$, phase resolution $\approx 0.1$ rad) and BHWD (right panel, SNR $\approx 3000$, phase resolution $\approx 0.02$ rad) cases.
    }\label{fig:case1000SNR}
\end{figure*}

Additionally, we conduct simulations with much higher SNRs, as the majority of systems exhibit a SNR around $\mathcal{O}(10^3)$ when considering a luminosity distance of $d_{\rm L}=20\rm$ kpc (comparable to the size of the Milky Way). Therefore, we conduct simulations with SNRs $\approx 1000$ and $3000$ for NSWD2 and BHWD cases, respectively, with the distance set to $d_{\rm L}=0.73\rm kpc$ for both cases. The simulation results are depicted in Fig.~\ref{fig:case1000SNR}, revealing that the nearly minimal achievable $\alpha$ values are $-9$ and $-10$, respectively, with corresponding phase resolutions of approximately $0.1$ rad and $0.02$ rad. The long tail at the lower end of $\alpha$ indicates that values in this region will provide an undetectable phase correction from MT effect. Although we have included currently available higher-order PN corrections to the total phase, as discussed in Appendices~\ref{subsec:degeneracy} and \ref{subsec:tidal_spin}, Fig. \ref{fig:case1000SNR} still shows a very strong degeneracy among $m_1$, $m_2$, and $\alpha$. In the future, additional corrections will be required to break this degeneracy.

\section{Conclusion and Discussion}\label{sec:conclusion}
In this work, to explore the detectability of MTR, we develop a waveform model that includes the MT effect and perform MCMC simulations to evaluate the measurement uncertainties on the MTR, $\dot{m}_2 \sim 10^\alpha \, M_\odot \, {\rm yr}^{-1}$, across various values and different scenarios. In both NSWD2 and BHWD cases, the approximate minimally detectable $\alpha$ values are $-7.1$ and $-7.7$, respectively, when the SNR is $20$. These values could be further reduced to $-9$ and $-10$ when SNR$\approx 1000$ and $3000$, respectively for NSWD2 and BHWD scenarios.\\

Our approach is independent of MT models, and GWs offer a direct method to measure the MTR. Assuming the MTR remains constant during the observing period in this study, in the future, to understand the evolution of MTR along the orbital frequency, conducting a systematic measurement of MTR at different frequency bins will result in a step distribution of the MTR, especially when the MTR varies in short timescales. Additionally, the spin of both the donor and accretor is expected to evolve along the orbital frequency due to mass transfer, presenting the potential for detection in the future. For our current analysis, we assume this spin remains constant. \\

A crucial aspect in the future will be to break the degeneracy among $m_1$, $m_2$, and $\alpha$, as illustrated in Fig.~\ref{fig:case1000SNR}. To achieve this, additional effects such as dynamical tides, modes, and others must be considered for both the phase and amplitude \cite{Fuller_2013,Piro_2019,McNeill_2019,Saio_2019,Perot_2022}. Furthermore, the motion of the detectors must be incorporated since the observation time usually spans years. \\

The successful measurement of MTR via GWs will not only advance the study of accretion mechanisms involving WD donors, but also assist in forecasting tidal disruption events or potentially supernova events with WD progenitors, thereby facilitating multi-messenger observations. \\


\begin{acknowledgments}
We thank Yong Gao, Yacheng Kang, Rui Xu for help discussions. This work is supported by the Beijing Municipal Natural Science Foundation (1242018), the National SKA Program of China (2020SKA0120300), the National Natural Science Foundation of China (11991053),  the Max Planck Partner Group Program funded by the Max Planck Society, and the High-performance Computing Platform of Peking University. Z.L. is supported by the Boya Postdoctoral Fellowship at Peking University.

\end{acknowledgments}

\software{Astropy \citep{astropy:2013, astropy:2018, astropy:2022}, Corner \citep{Mackey_2016}, Dynesty \citep{Speagle_2020}, Matplotlib \citep{Hunter:2007}, Numpy \citep{numpy_5725236}, PyCBC \citep{Biwer:2018osg}, Scipy \citep{scipy_4160250} }

\appendix

\section{Continuous Mass Transfer}\label{sec:continuous_MT}
UCXBs are classified as either persistent or transient sources, depending on the mass-transfer rate $\dot{m}_2$ of the system, and further, on the orbital period \citep{Nelemans_2010,Sengar_2017}. Driven by the loss of angular momentum due to GW radiation, as the system approaches tidal disruption in this study, we assert that there will be a continuous mass flow through the disk to maintain the ionization of the entire accretion disk \citep{LASOTA2001449,Heinke_2013,Sengar_2017}. \\

The initial orbital frequency at which MT initiates is determined by the point where the radius of the WD fills its effective Roche Lobe. The well-known fitting formula for the Roche Lobe radius $r_{\rm L}$, accurate to $1\%$ for all mass ratios, is provided by \citep{Eggleton_1983}
\begin{align}
    \frac{r_{\rm\, L}}{r} = \frac{0.49\, q^{2/3}}{0.6\, q^{2/3}+\ln{(1+q^{1/3})}}\,,
\end{align}
Here, $q=m_2/m_1$ represents the ratio of the donor to the accretor, and $r$ is the orbital separation of the two objects. Applying Kepler's third law $\Omega^2\,r^3=G(m_1+m_2)$, when the Roche Lobe radius reaches the WD radius $r_{\rm\, L}=R_2$, the orbital angular frequency $\Omega$ at which MT takes place can be determined by solving the above equations, considering the universal relation between WD radii and masses as well \citep{Tout_1997,Carvalho_2015}. \\

To ensure that WDs can withstand the tidal effect caused by the NS or BH, there is a termination orbital separation where tidal disruption occurs. The tidal disruption radius is estimated as follows \citep{Maguire_2020}
\begin{align}
    r_{\rm tidal} = R_2\left(\frac{m_1}{m_2}\right)^{1/3}
\end{align}
Here, $R_2$ and $m_2$ represent the radius and mass of the donor (WD), and $m_1$ is the mass of the accretor (NS/BH). Utilizing Kepler's third law again, we can determine the orbital frequency at tidal disruption. We have computed the GW frequency (twice of the orbital frequency) associated with the initiation of  Roche Lobe overflow and the tidal disruption, as presented in column 4 and 6 of Table~\ref{tab:cases}. The duration of MT decreases dramatically with the decreasing radius of the WD. This phenomenon may partially account for why the donors in observed  UCXBs have a mass around $0.05 \rm M_\odot$, although this estimation relies on the specific model employed \citep{yungelson2002formation,Heinke_2013,Sengar_2017}.

We neglect the influence of the accretion disk, assuming it to have an axially symmetric distribution and treating it as an integral part of the rotating accretor \citep{Wilkins_2017,wang_2016}. Furthermore, in the Markov Chain Monte Carlo (MCMC) simulation in Sec.~\ref{sec:pe}, the spin of the accretor is set to $\chi_1=0.05$~\citep{Kiziltan_2013}, given that significant mass accretion would result in the spin-up of the NS. \\

\section{Waveform Model}\label{sec:waveform_model}
For a binary system with masses $m_1$ and $m_2$ where $ m_1 \geq m_2$, we introduce the following parameters
\begin{align}
    M = m_1 + m_2\,,\quad X_1 = m_1/M\,,\quad
    X_2 = m_2/M\,,\quad
    \eta = X_1 X_2\,,\quad 
    \mathcal{M}_c = \eta^{3/5} M\,,
\end{align}
 
 The Post-Newtonian (PN) expansion parameter is defined as
\begin{align}\label{eq:v}
    v:=\left(\frac{GM}{c^3} \frac{d\phi}{dt}\right)^{1/3} = \left(\frac{GM}{c^3} \pi\, f\right)^{1/3}\,,
\end{align}
where $\phi(t)$ is the orbital phase, and $f$ is the instantaneous GW frequency (twice the orbital frequency). For a typical binary system involving WDs, the invariant velocity is
\begin{align}
    v= 0.014\left(\frac{M}{2\rm M_\odot}\right)^{1/3} \left(\frac{f}{0.1\rm Hz}\right)^{1/3} << 1\,,
\end{align}
Therefore, expressing all evolving quantities in terms of $v$ is appropriate.\\

\subsection{Mass Transfer Induced Phase}\label{subsec:MT_phase}
Assuming the MT rate (MTR) of the donor is $\dot{m}_2$ (the dot denotes time derivatives, and $\dot{m}_2<0$), and the fraction of mass loss from the donor that is transferred to the accretor is $\dot{m}_1=-\beta\,\dot{m}_2$ with $0\leq \beta \leq1$. The total mass loss from the system will be $\dot{M}=(1-\beta)\,\dot{m}_2\leq0$. When $\beta=1$, the system maintains total mass conservation, whereas for  $\beta<1$, the binary system undergoes a gradual loss of total mass. Due to the MT effect, the adiabatic approximation is no longer valid. We must employ the modified balance equation, which is
\begin{align}\label{eq:balance}
    \frac{d E(v;m_i(v))}{d t} + \mathcal{F}(v;m_i(v)) + \dot{M} = 0\,,
\end{align}
where $m_i(v)$ is the mass of the $i$th object. $E(v;m_i(v))$ and $\mathcal{F}(v;m_i(v))$ are the orbital energy and the GW flux to infinity, respectively. With the chain rule, $d E / d t=E'(v;m_i(v))\,dv/dt$, we obtain
\begin{align}\label{eq:dtdv}
    \frac{dt}{dv} = - \frac{E'(v;m_i(v))+\dot{M}\,(dt/dv)^{\rm I}}{\mathcal{F}(v)}\,,
\end{align}
where $(dt/dv)^{\rm I}$ is derived from the initial value of masses $m_i^{\rm I}$. Using the definition of $v$ in Eqn.~\ref{eq:v}, we see
\begin{align}\label{eq:dphidv}
    \frac{d\phi}{dv} = \frac{v^3}{M(v)}\,\frac{dt}{dv} = -\frac{v^3}{M(v)}\,\frac{E'(v;m_i(v))+\dot{M}\,(dt/dv)^{\rm I}}{\mathcal{F}(v)}\,,
\end{align}

Given the expressions of $E(v)$ and $\mathcal{F}(v)$ and after re-expanding them in a Taylor series, the equations above can be integrated to find $t(v)$ and $\phi(v)$. The phase of the TaylorF2 waveform model then reads
\begin{align}\label{eq:psi_f}
    \psi(f) = 2\pi f\, t(v) -2\phi(v) - \frac{\pi}{4}\,,
\end{align}\\

To clearly distinguish the effects from GW radiation and MT, it is better to separate the phases of each effect. To achieve this, the expressions of \(E(v)\) and \(\mathcal{F}(v)\) can be split into two terms
\begin{align}
    E(v;m_i(v)) = E^{\rm I}(v) + \delta E(v)\,,\quad
    \mathcal{F}(v;m_i(v)) = \mathcal{F}^{\rm I}(v) + \delta \mathcal{F}(v)\,,
\end{align}
where $E^{\rm I}(v)$ and $\mathcal{F}^{\rm I}(v)$ are defined in terms of initial value of masses $m^{\rm I}_i$. $\delta E(v)$ and $\delta \mathcal{F}(v)$ are introduced by $\delta m_i(v)$. Here, we consider $\delta m_i(v)$ as a small quantity (typically much smaller than $10^{-2}\rm M_\odot/yr$), and it's straightforward to re-expand $E$ and $\mathcal{F}$ in power of $v$ by replacing\footnote{We have assumed $\delta m_i(v)=\delta m_i(v^{\rm I})$}
\begin{align}
    m_i(v) &\longrightarrow m_i^{\rm I}+\delta m_i(v)\,,\\
    v &\longrightarrow(\pi M(v) f)^{1/3} = (\pi M^{\rm I} f)^{1/3}\,(1+\frac{\delta M}{M^{\rm I}})^{1/3} = v^{\rm I} + \frac{1}{3}\frac{\delta M}{M^{\rm I}}\,v^{\rm I}\,,
\end{align}
and keep the leading order of MT effect. The leading order expressions are
\begin{align}
    E^{\rm I}(v) &= -\frac{M\eta}{2}\,v^2\,,\quad
    \mathcal{F}^{\rm I}(v) = \frac{32}{5} \eta^2 v^{10}\,,\\
    \delta E(v) &= -\frac{1}{6}\,\xi\,\delta m_2\, v^2\,,\quad
    \delta \mathcal{F}(v) = \frac{64\,\eta}{15M}\,\xi\,\delta m_2\,v^{10}\,,
\end{align}
where $\xi=3X_1 - 3\,\beta\,X_2 + \beta\,\eta -\eta$. After plugging $E(v;m_i(v))$ and $\mathcal{F}(v;m_i(v))$ into Eqn.~\ref{eq:dtdv} and Eqn.~\ref{eq:dphidv}, we have
\begin{align}
    t(v) = t^{\rm I}(v) + \delta t(v)\,,\quad
    \phi(v) = \phi^{\rm I}(v) + \delta \phi(v)\,,
\end{align}
where
\begin{align}
     t^{\rm I}(v) &=  -\frac{5M}{256\,\eta\, v^8}\,,\quad
     \delta t(v) = \frac{5\,\xi\,\delta m_2}{768\,\eta^2\,v^8} + \frac{25(1-\beta)M \dot{m}_2}{18432\,\eta^3\,v^{18}} \,,\\
     \phi^{\rm I}(v) &= -\frac{1}{32\,\eta\,v^5}\,,\quad
     \delta \phi(v) = -\frac{\xi\,\delta m_2}{96\,M\,\eta^2\,v^5} + \frac{5(1-\beta)\dot{m}_2}{3072\,\eta^3\,v^{15}} \,,
\end{align}
After substituting above expressions into Eqn.~\ref{eq:psi_f}, we obtain
\begin{align}
    \psi_{\rm pp} &= 2\pi f t_c - \phi_c -\frac{\pi}{4} + \psi_{\rm 0PN}\left[1+\mathcal{O}(v^2)\right]\,,\quad
    \psi_{\rm 0PN} = \frac{3}{128\,\eta\,v^5}\,,\\ \label{eq:psi_pp}
    \psi_{\rm MT} &= -\frac{\xi\,\delta m_2}{128\,M\,\eta^2\,v^5} - \frac{5(1-\beta) \dot{m}_2}{9216\,\eta^3\,v^{15}} \,,
\end{align}
where $\psi_{\rm pp}$ and $\psi_{\rm MT}$ are the leading PN order terms. Regarding the MT term, $\delta m_2 = \dot{m}_2\,\delta t$, with the leading order of $\delta t$ from GW radiation being
\begin{align}
    \delta t^{\rm I}(v) = \frac{5M}{32\,\eta\,v^9}\,\delta v^{\rm I}\,,
\end{align}
Finally, we have
\begin{align}
    \psi_{\rm MT} &= -\frac{5\,\xi\,\dot{m}_2\,\delta v^{\rm I}}{4096\,\eta^3\,v^{14}} - \frac{5(1-\beta)\,\dot{m}_2}{9216\,\eta^3\,v^{15}}\,,\label{eq:phiMT2}
\end{align}

As the waveform is in frequency domain, $\delta v^{\rm I}=v-v_i$ where $v_i$ is the initial value of $v$. The first term is proportional to $v^{-13}=v^{-5}\,v^{-8}$ which is -4PN order. The second term comes from the non-conservative of the total mass, which is -5PN. In other words, when the total mass of the system is conserved ($\beta=1$), the second term vanishes.\\

\subsection{Degeneracy Analysis}\label{subsec:degeneracy}
The first degeneracy arises in the $-5$PN term of MT phase in Eqn.~\ref{eq:phiMT2}. It is only feasible to estimate the product $(1-\beta)\,\dot{m}_2$ as a whole, making it impossible to simultaneously estimate both $\beta$ and $\dot{m}_2$ from GWs. Certainly, if there is a constraint on either $\dot{m}_2$ or  $\beta$  from electromagnetic observations (likely dependent on the MT model), GWs could assist in providing uncertainty estimates for the other.

When the total mass is conserved ($\beta=1$), the degeneracy is primarily induced by the monochromatic feature of the orbital evolution. A complete degeneracy exists between $M$ and $f$ in the PN velocity $v=(\pi M f)^{1/3}$. In principle, integrating higher PN order terms to point particle phase $\psi_{\rm pp}$  and additional corrections into the overall phase could potentially address the mass-frequency degeneracy. We have incorporated currently available corrections from point particle radiation up to $3.5$PN order \citep{GW_power_1963,Arun_2005,Buonanno_2009}, tidal deformation up to $7.5$PN \citep{Blanchet_1995,Vines_2011,Damour_2012,Agathos_2015}, and spin-induced phase up to $3.5$PN order \citep{Blanchet_2006}. In Sec.~\ref{subsec:tidal_spin} , we investigate the impact of phase corrections arising from tidal and spin effects. To enhance the accuracy of the waveform model, we can incorporate corrections from dynamic tidal or other important modes in the future, and hence, these adjustments can also contribute to breaking the existing degeneracy.\\

\subsection{Tidal and Spin Effects}\label{subsec:tidal_spin}
Tidal deformations, represented by the mass-quadrupole moment $Q_{ij}$ (specifically of WDs in this study), are induced by the tidal tensor $\mathcal{E}_{ij}$, which originates from their companion. In the adiabatic approximation, these two quantities are related by
\begin{align}
    Q_{ij} = -\lambda(m)\,\mathcal{E}_{ij}\,,\quad
    \lambda(m) := \frac{2}{3}\,k_2\,\frac{R^5}{G}\,,
\end{align}
 the coefficient $\lambda(m)$ represents the tidal deformability. Here, $k_2$, $m$ and $R$ denote the second Love number, mass and radius of the object undergoing the quadrupole deformation, respectively. The tidal deformability $\lambda(m)$ is determined by the equation of state. The deformations experienced by the WDs, in turn, accelerate the orbital evolution, thereby leaving a signature on the gravitational waveform. We will incorporate phase corrections up to $7.5$PN \citep{Blanchet_1995,Vines_2011,Damour_2012,Agathos_2015} with the leading order effect occurring at $5$PN
\begin{align}\label{eq:tide}
    \psi_{\rm tide} = \psi_{\rm 0PN}\sum_{i=1,2}\left[(-288+264X_i)X_i^4\,\Lambda_i\,v^{10} + \mathcal{O}(v^{12})\right]\,,\quad
    \Lambda := \frac{2}{3}\,k_2\,\left(\frac{c^2R}{G m}\right)^5
\end{align}
where $\Lambda_i$ is the dimensionless tidal deformability, and $k_2$ is approximately in the range of 0.05 to 0.15 for both WDs \citep{Taylor_2019,Perot_2022} and realistic neutron stars \citep{Hinderer_2008,Damour_2009,Binnington_2009}, and for black holes, $k_2=0$ \citep{Damour_2009,Binnington_2009}. 
Nevertheless, the radius of WDs is roughly 1000 times that of neutron stars, leading to an exceedingly larger dimensionless parameter $\Lambda $ ($\approx 10^{15}-10^{22}$) for WDs. Fortunately, the inspiraling velocity of binaries involving WDs is much smaller than the late-inspiral stage of NS-BH binaries. This ensures that the $5$PN coefficient remains of the same order of magnitude between coalescing WD-NS and NS-BH binaries, allowing the Taylor expansion to remain valid.

TaylorF2 is an aligned-spin waveform model, where the spins of component masses align with the orbital angular momentum. The leading order effect from spin-induced phase is given by
\begin{align}
    &\psi_{\rm spin} = \psi_{\rm 0PN}\bigg[c_{\rm s1s2} + \sum_{i=1,2}\big(c_{\rm SO} + c_{\rm self-spin} + c_{\rm QM}\big)\bigg]\,,\\
    &c_{\rm SO} = \Big(25+\frac{38}{3}X_i \Big)\,X_i\,(\vec{\chi}_i\cdot\hat{L})\, v^3 + \mathcal{O}(v^4)\,,\\ 
    &c_{\rm s1s2} = \frac{5}{24}\bigg(247(\vec{\chi}_1\cdot\vec{\chi}_2) -721(\hat{L}\cdot\vec{\chi}_1)(\hat{L}\cdot\vec{\chi}_2)\bigg)\,v^4
    + \mathcal{O}(v^6) \,,\\
    &c_{\rm self-spin} = -\frac{35}{48}\,X_i^2\,\chi_i^2\, v^4 + \mathcal{O}(v^6) \,,\\
    &c_{\rm QM} =-25\kappa_i\,\left(3(\hat{\chi}_i\cdot\hat{L})^2-1\right)X_i^2\,\chi_i^2\, v^4 + \mathcal{O}(v^6) \,, 
\end{align}
where $\hat{\chi}_i$ represents the unit vector along the spin direction of the $i$th object and $\hat{L}$ is the unit vector along the direction of the orbital angular momentum. Additionally, $\chi_i$ and $\kappa_i$ are the dimensionless spin magnitude and spin-induced quadrupole-moment coefficient, respectively, as defined in Eqn.~\ref{eq:chi} and Eqn.~\ref{eq:kappa}. As for the values of $\Lambda_i$, $\chi_i$, and $\kappa_i$, there are well-established universal relations involving the mass of WDs discussed in Sec.~\ref{subsec:universal}. \\

The dominant spin effect arises from the spin-orbit coupling $\psi_{\rm SO}$ at $1.5$PN. In the case of a spin-orbit synchronized NS-WD binary with $m_1=1.4\rm M_\odot$ and $m_2=0.5\rm M_\odot$, the spin $\chi_2\approx 20$ at $0.1$ Hz. The coefficient $1.5$PN order effect, given by 
\begin{align}
    X_i\left(25+ \frac{38}{3} X_i \right)\,\chi_i\, v^3\approx 5\times 10^{-4} \ll 1
\end{align}\\

\subsection{Universal Relations}\label{subsec:universal}
To reduce intrinsic parameters in MCMC simulations, we employ three well-known universal relations to eliminate dimensionless quantities: $\Lambda$, $\chi$, and $\kappa$ from the waveform (ignore the subscript $i$ in this section). For cold, slowly rotating WDs, We obtain $\bar{I}$, $\Lambda$, and $\bar{Q}$ from universal relations \citep{Wolz_2020,Tang_2023,Yagi_2013,Boshkayev_2016,Taylor_2019,Roy_2021}
\begin{align}
    \nonumber\ln{\bar{I}} &= 24.7995 - 39.0476 m_{\rm 1M_\odot} + 95.9545 m_{\rm 1M_\odot}^2 - 138.625 m_{\rm 1M_\odot}^3 + 
 98.8597 m_{\rm 1M_\odot}^4 - 27.4000 m_{\rm 1M_\odot}^5\,,\\
     \ln{\Lambda} &= 2.02942 + 2.48377 \ln{\bar{I}}\,,\\
     \ln{\bar{Q}} &= 1.89977 + 0.48372 \ln{\bar{I}} \,,
\end{align}
where $m_{\rm 1M_\odot}$ is the mass of a WD in unit of solar mass. The relative error of the relations is less than $\sim 1\%$ \citep{Wolz_2020}. \\

According to $\bar{I}$, we have the dimensionless spin parameter 
\begin{align}\label{eq:chi}
    \chi := \frac{J}{m^2} \frac{c}{G} = \frac{I \Omega}{m^2} \frac{c}{G} = \bar{I}\Omega m \frac{G}{c^3}\,,
\end{align}
where the angular momentum $J=I\Omega$, and $\Omega$ is the orbital angular frequency. Moreover, spin-orbit synchronization is achieved before to MT or tidal disruption, as demonstrated in the models presented in paper \citep{Fuller_2012}. Simultaneously, the system approaches its maximum angular velocity, calculated from the Keplerian angular velocity \citep{Boshkayev_2013, Boshkayev_2016}. As a result, we can express $\Omega$ as $\pi f$ where $f$ denotes the gravitational wave frequency.\\

The dimensionless rotation-induced quadrupole moment $\bar{Q}$, which is equivalent to the spin-induced quadrupole moment $a$ in Ref.~\citep{Poisson_1998}, is denoted as $\kappa$ in GW waveform modeling
\begin{align}\label{eq:kappa}
    \kappa  = - \frac{Q^{(\rm rot)}}{m^3\chi^2} = -\frac{Q^{(\rm rot)}}{J^2/m } = \bar{Q} \,,
\end{align}\\

\section{Bayesian Inference}\label{subsec:Bayes}
\begin{figure*}[th]
    \centering
    \includegraphics[width=0.9\linewidth]{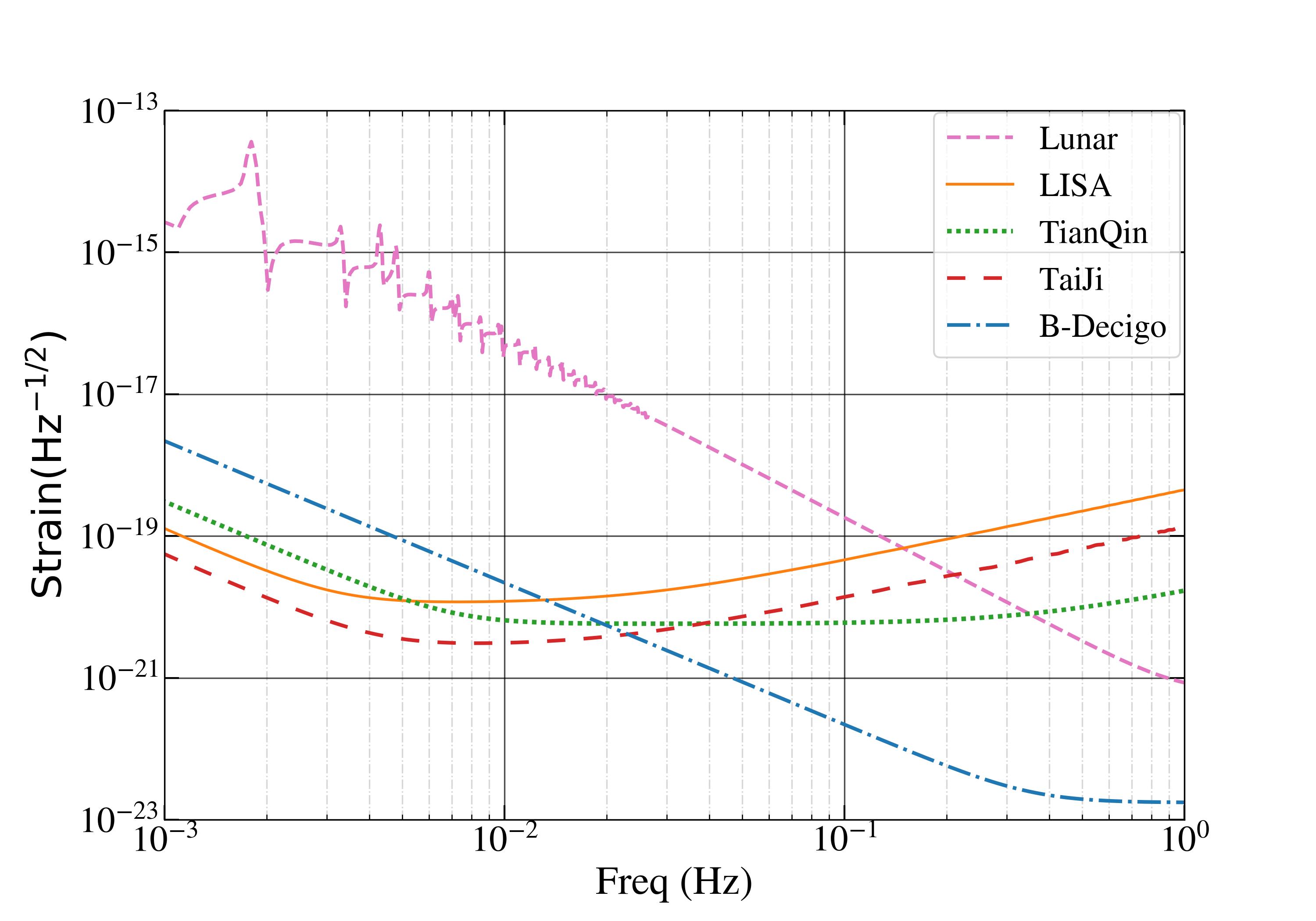}
    \caption{Sensitivity curves of various detectors. The Lunar sensitivity is based on \citet{Junlang_2023}. The LISA, TianQin, TaiJi and B-Decigo sensitivities are the sky position and GW polarization averaged noise spectral density, taken from \citet{babak2021lisa,Robson_2019,Isoyama_2018,Liu_2023,Mei_2020}.}
    \label{fig:sensitivity}
\end{figure*}

We will employ Bayesian inference to estimate the measurement uncertainty of the MTR. This approach is built upon MCMC and Bayes' theorem \citep{Flanagan_1998,Bayes2019, Bayes2021}
\begin{equation}
p(\mathcal{\boldsymbol{\vartheta}} | d) =  \frac{p(d | \boldsymbol{\vartheta} )\, p(\boldsymbol{\vartheta})}{p(d)} \,,
\end{equation}
here $p(\boldsymbol{\vartheta})$ denotes the prior probability distribution on $\mathcal{\boldsymbol{\vartheta}}= (\mathcal{M}_c, \eta, \alpha, d_{\rm L}, \iota, t_c, \phi_c)$, representing the set of parameters to be estimated. These parameters are the chirp mass, symmetric mass ratio, power index of MTR, luminosity distance, inclination angle, coalescence time, and coalescence phase, respectively. The posterior distribution of $\boldsymbol{\vartheta}$ given the data $d$ is denoted as $p(\boldsymbol{\vartheta} | d)$, with $p(d)$ acting as the evidence and treated as a normalization factor. Assuming stationary Gaussian noise, the log-likelihood can be expressed as
\begin{equation}
  \log p(d|\boldsymbol{\vartheta}) \propto -\frac{1}{2} \langle d - h(\boldsymbol{\vartheta}) \,, d - h(\boldsymbol{\vartheta}) \rangle\,,
\end{equation}
where $d$ and $h(\boldsymbol{\vartheta})$ are the data and waveform template, respectively. The noise-weighted inner product, denoted by $\langle\cdot\,,\cdot\rangle$, is \citep{Finn_1992, babak2021lisa}
\begin{align}
    \langle a(t)\,, b(t) \rangle = 4\,{\rm Re}\int_{f_{\rm i}}^{f_{\rm t}} \frac{\tilde{a}(f)\, \tilde{b}^*(f)}{S_n(f)} df\,, 
\end{align}
where $*$ denotes a complex conjugate. The variables $f_{\rm i}$ and $f_{\rm t}$ represent the initial frequency and the tidal disruption frequency, respectively, as indicated in Table~\ref{tab:cases}..  $S_n(f)$ represents the one-sided noise power spectral density (PSD) of given detectors, averaged over sky position and GW polarization \citep{Robson_2019,babak2021lisa}, as illustrated in Fig.~\ref{fig:sensitivity}.

We adopt five different detectors to estimate the optimal matched filtering SNRs of the signal $\tilde{h}(f)=F_+\,\tilde{h}_+(f) + F_\times\,\tilde{h}_\times(f)$
\begin{align}
    {\rm SNR}^2 = 4\int_{f_{\rm i}}^{f_{\rm t}} \frac{|\tilde{h}_+(f)|^2 + |\tilde{h}_\times(f)|^2}{S_n(f)} df\,, 
\end{align}
where $F_+$ and $F_\times$ are the frequency dependent detector response functions. These functions depend on the sky location and polarization angle of the source and have already been incorporated into the averaged PSD. The SNRs for the considered scenarios are provided in columns 8 to 12 of Table~\ref{tab:cases}.

\clearpage
\bibliography{ref.bib}{}

\end{document}